\newcommand {\NP}   {Nucl.Phys.}
\newcommand {\PL}   {Phys.Lett.}
\newcommand {\PR}   {Phys.Rev.}
\newcommand {\PRL}   {Phys.Rev.Lett.}

\documentclass{article} 
\usepackage{amssymb} 
\usepackage{amsfonts} 
\begin{document}

\begin{flushright}
EDO-EP-34\\
September, 2000\\
\end{flushright}
\vspace{30pt}

\begin{center}
       \textbf{Localization of Various Bulk Fields on a Brane}\footnote{
          The expanded version of an invited paper which appears in
          "Concise Encyclopedia on Supersymmetry
           and Noncommutative Structures in Mathematics and Physics",
           edited by J. Bagger, S. Duplij and W. Siegel, Kluwer Academic 
           Publishers, Dordrecht.}

\vspace{5mm}

Ichiro Oda\footnote{E-mail: ioda@edogawa-u.ac.jp
          }

\vspace{3mm}

Edogawa University, 474 Komaki, Nagareyama City, Chiba 270-0198, JAPAN

\end{center}


\vspace{7mm}
\begin{abstract}
We report the issues of localization of various bulk fields on a
brane in a curved background from a local field-theoretic viewpoint 
where a special attention is paid to a warp geometry in a general space-time 
dimension.
We point out that spin 0 scalar field and spin 2 graviton are naturally
localized on a brane with the exponentially decreasing warp factor, while
spin 1/2 and 3/2 fermionic fields can be localized on the brane by 
introducing a bulk mass term with a 'kink' profile.
{}For the localization of spin 1 vector field on the brane with the 
exponentially decreasing warp factor, it is shown that there are essentially
two ways. One way is to appeal to the Dvali-Shifman method which is based on
a peculiar feature of the non-abelian gauge dynamics such as bulk
confinement, 
whereas the other way is to look for a solution of Einstein equations in such
a way that the geometry has a property of confining the vector field to the 
brane by a gravitational interaction as in six dimensions. We point out
that the
latter way is more universal.

\vspace{10mm}

\end{abstract}

It is now widely believed that extra dimensions play an important
role in constructing a unified theory of all interactions, such as
superstring theory. 
In recent years, the idea of a brane world in a curved space-time 
\cite{Rubakov, Akama, Visser, Randall1, Randall2, Oda1, Oda2} 
has been put forward as explaining the extremely small value of the 
cosmological constant, 
the huge mass hierarchy between the Planck scale and the electroweak scale and
a possible source of the supersymmetry breaking and so on.

In the brane world scenario, our universe is regarded as a 3-brane embedded 
in a higher-dimensional space-time. 
Then a key ingredient for realizing the brane world idea is localization of 
various bulk fields on a brane by a natural mechanism. In other words, in
this 
scenario various fields we observe in our universe are nothing but the zero 
modes of the corresponding bulk fields which are trapped on our brane by some 
ingenious mechanism. This mechanism is certainly provided by the
nonperturvative
effects in M-theory \cite{Horava} and D-brane theory \cite{Polchinski}, 
but it is at present far from complete 
to understand the precise procedure of getting a desired model of the brane
world 
from M-theory and D-brane theory even if there have been many efforts and
attempts so far.

Recently, a complete understanding of the localization
of various spin fields has been reached within the framework of a local
field theory, so the aim of this review article is to present the
localization 
mechanism in a concise manner. Even if there might appear the other
localization
mechanisms in future, our localization mechanism is quite beautiful and
universal in that only the universal interaction, that is, the gravitational
interaction, is utilized for the trapping of bulk fields on a brane (in six
space-time dimensions) so we think the essential idea behind our localization
mechanism would remain true. 

Let us fix our physical setup. We shall consider $D = D_1 + D_2 +
1$-dimensional 
geometry with the line element
\begin{eqnarray}
ds^2 = e^{-A(r)}\hat{g}_{\mu\nu}(x^\lambda) dx^\mu dx^\nu + dr^2
+ e^{-B(r)}\tilde{g}_{mn}(y^l) dy^m dy^n,
\label{1}
\end{eqnarray}
where $\mu, \nu, \lambda = 0, 1, \cdots, D_1-1$ and $m, n, l 
= 1, 2, \cdots, D_2$.
The coordinates $x^\mu$ and $y^m$ are respectively the ones on the brane
and the extra internal manifold, and the coordinate $r$ runs over
from $-\infty$ to $+\infty$. We have the physical situation in mind that
a ($D_1-1$)-brane is supposed to sit at the origin of the coordinate $r$
and then ask whether or not various bulk fields with spin ranging from 0 to 2 
can be localized on the brane in terms of only the gravitational interaction.
If the zero modes of all the bulk fields are trapped on the brane, it is
reasonable to regard the brane as a candidate of our world equipped with
necessary physical properties. Note that this way of thought provides us
an alternative possibility of compactification where the size of internal
manifold is not always fixed to be order of the Planck scale. In what follows,
we shall study the property of localization of bulk fields according to 
their spin in order. In this article, we will assume that the background
metric is not modified by the presence of the bulk fields, namely, we will
neglect the back-reaction on the metric from the bulk fields. 

First of all, we shall consider localization of a real scalar field 
on a ($D_1-1$)-brane in the above background geometry. 
The action takes the form
\begin{eqnarray}
S_0 = -\frac{1}{2} \int d^Dx \sqrt{-g} g^{MN} \partial_M \Phi
\partial_N \Phi,
\label{2}
\end{eqnarray}
where $M, N$ denote $D$-dimensional space-time indices. From the equation
of motion stemming from this action, $\partial_M(\sqrt{-g}g^{MN}\partial_N
\Phi) = 0$, we can find the massless zero-mode solution with the form of 
$\Phi(x^M) = \phi(x^\mu) u_0 \chi(y^m)$ where $\partial_\mu
(\sqrt{-\hat{g}}\hat{g}^{\mu\nu} \partial_\nu) \phi = \partial_m
(\sqrt{\tilde{g}}\tilde{g}^{mn} \partial_n) \chi = 0$ and $u_0$ is
a constant. Substituting this solution into the starting action, we
obtain
\begin{eqnarray}
S_0 = -\frac{u_0^2}{2} \int d^{D_2}y \sqrt{\tilde{g}} \chi^2(y)
\int_{-\infty}^\infty dr e^{\frac{2-D_1}{2}A - \frac{D_2}{2}B}
\int d^{D_1}x \sqrt{-\hat{g}} \hat{g}^{\mu\nu} \partial_\mu \phi
\partial_\nu \phi. 
\label{3}
\end{eqnarray}
The condition that the zero-mode solution is localized on a brane
is equivalent to the normalizability of the ground state wave function
on the brane \cite{Bajc}.
Then, with the assumption that the first integral over $y$ is finite,
the condition requires the second integral over $r$ to be finite.
In particular, when we consider a warp geometry $A = c_1 |r|, B = c_2$
with positive constants $c_1$ and $c_2$ which is indeed known to be a 
nonsingular solution to Einstein's equations in an arbitrary dimension
except in six dimensions \cite{Vilenkin, Oda3}, 
this condition reduces to $\frac{2-D_1}{2}c_1 < 0$. 
Note that this inequality is always satisfied when $D_1 =4$, thereby
making a spin 0 bulk scalar field trap on our universe \cite{Bajc}.

Next let us turn our attention to spin 1/2 spinor field. The starting action
is now given by
\begin{eqnarray}
S_{1/2} = \int d^Dx \sqrt{-g} \bar{\Psi} i (\Gamma^M D_M
+ m \varepsilon(r)) \Psi, 
\label{4}
\end{eqnarray}
where the covariant derivative is defined as $D_M \Psi = ( \partial_M + 
\frac{1}{4} \omega_M^{AB} \gamma_{AB} ) \Psi$ with the definition of
$\gamma_{AB} = \frac{1}{2} [\gamma_A, \gamma_B]$, and $\varepsilon(r)$
is $\varepsilon(r) \equiv \frac{r}{|r|}$ and $\varepsilon(0) \equiv 0$.
Here the indices
$A, B$ are the ones of the local Lorentz frame and the gamma matrices
$\Gamma^M$ and $\gamma^A$ are related by the vielbeins $e_A^M$ 
through the usual relations $\Gamma^M = e_A^M \gamma^A$ where 
$\{\Gamma^M, \Gamma^N\} = 2 g^{MN}$.
An important feature of the action is the existence of mass
term with a 'kink' profile. (It is plausible to consider that such
a mass term is generated by the Higgs field with a vacuum expectation
value of a 'kink' profile \cite{Jackiw} 
but we do not argue the origin of the term at
present.)  
In the background (\ref{1}), the spin connections are easily evaluated
to be
\begin{eqnarray}
\omega_\mu = \frac{1}{4} A' \Gamma_r \Gamma_\mu + 
\hat{\omega}_\mu(\hat{e}), \ \omega_r = 0, \
\omega_m = \frac{1}{4} B' \Gamma_r \Gamma_m + 
\tilde{\omega}_m(\tilde{e}), 
\label{5}
\end{eqnarray}
where the prime denotes the differentiation with respect to the coordinate
$r$ and we have defined $\omega_M \equiv \frac{1}{4} \omega_M^{AB}
\gamma_{AB}$.
Using Eq. (\ref{5}), the Dirac equation $(\Gamma^M D_M + m \varepsilon(r)) 
\Psi = 0$ can be cast into the form
\begin{eqnarray}
\left[\Gamma^r (\partial_r - \frac{D_1}{4} A' - \frac{D_2}{4} B') 
+ \Gamma^\mu (\partial_\mu + \hat{\omega}_\mu) + \Gamma^m (\partial_m + 
\tilde{\omega}_m) + m \varepsilon(r) \right] \Psi = 0. 
\label{6}
\end{eqnarray}
Let us find the massless zero-mode solution with the form of $\Psi(x^M) = 
\psi(x^\mu) u(r) \chi(y^m)$ where $\Gamma^\mu \hat{D}_\mu \psi = 
\Gamma^m \tilde{D}_m \chi = 0$ and the chirality condition $\Gamma^r \psi =
\psi$ 
is imposed on the brane fermion. Then $u(r)$ is solved to be 
$u(r) = u_0 e^{\frac{D_1}{4}A + \frac{D_2}{4}B - m \varepsilon(r) r}$ with
a constant $u_0$. Plugging this solution into the starting action
(\ref{4}), the action reduces to the form
\begin{eqnarray}
S_{1/2} = u_0^2 \int d^{D_2}y \sqrt{\tilde{g}} \chi^\dagger(y)\chi(y)
\int_{-\infty}^\infty dr e^{\frac{1}{2}A - 2 m \varepsilon(r) r}
\int d^{D_1}x \sqrt{-\hat{g}} \bar{\psi} i \gamma^\mu \hat{D}_\mu
\psi, 
\label{7}
\end{eqnarray}
where we have used $\Gamma^\mu = e_a^\mu \gamma^a = e^{\frac{1}{2}A}
\gamma^\mu$. Again the condition of the trapping of the bulk spinor
on a brane requires that the integral over $r$ has a finite value.
If we consider a warp geometry with $A = c_1 |r|$ with some positive
constant $c_1$, the condition becomes $\frac{1}{2}c_1 - 2m < 0$. 
Note that this inequality holds
as long as the mass is large enough compared to $c_1$. 
It is worthwhile to notice that if the bulk mass vanishes ($m = 0$), 
the zero mode of the bulk fermion cannot be localized on the brane.
This mechanism of localization of spin 1/2 fermions on a brane was
first discovered in a flat space-time long ago by Jackiw and Rebbi 
\cite{Jackiw} and recently extended to the case of $AdS_5$ by Grossman 
and Neubert \cite{Grossman} and more recently applied to the case of higher
dimensional space-times by Randjbar-Daemi and Shaposhnikov \cite{Randjbar}.

Now we are ready to consider spin 1 $U(1)$ vector field. (The generalization
to the nonabelian gauge fields is straightforward. And the inclusion of
bulk mass does not change the results obtained below.) The action reads
\begin{eqnarray}
S_1 = -\frac{1}{4} \int d^Dx \sqrt{-g} g^{MN} g^{RS} F_{MR} F_{NS}, 
\label{8}
\end{eqnarray}
where $F_{MN} = \partial_M A_N - \partial_N A_M$. The equations of motion
$\partial_M (\sqrt{-g} g^{MN} g^{RS} F_{NR}) = 0$ can be solved with
the ansatz of $A_r = A_m = 0$. (This ansatz is indeed justified since
the Kaluza-Klein modes associated with $A_r$ and $A_m$ have an infinite
energy so they are unphysical \cite{Oda4}.) The massless zero-mode solution
is of the form $A_\mu(x^M) = a_\mu(x^\mu) u_0 \chi(y^m)$ where 
$\hat{\nabla}^\mu a_\mu = \partial^\mu f_{\mu\nu} = 0$ with
$f_{\mu\nu} \equiv \partial_\mu a_\nu - \partial_\nu a_\mu$.
The substitution of this solution into the action leads to
\begin{eqnarray}
S_1 = -\frac{u_0^2}{4} \int d^{D_2}y \sqrt{\tilde{g}} \chi^2(y)
\int_{-\infty}^\infty dr e^{\frac{4-D_1}{2}A - \frac{D_2}{2}B}
\int d^{D_1}x \sqrt{-\hat{g}} \hat{g}^{\mu\nu} \hat{g}^{\lambda\sigma}
f_{\mu\lambda}f_{\nu\sigma}. 
\label{9}
\end{eqnarray}
The localization condition of this zero mode on a brane requires
the integral over $r$ to be finite. Localizing the massless vector
field is a subtle issue, so here let us discuss it in detail. 
Provided that we consider a warp geometry $A = c_1 |r|, B = c_2$
with positive constants $c_1$ and $c_2$ as in a scalar field, 
this condition reduces to $\frac{4-D_1}{2}c_1 < 0$, so in $D_1 = 4$
the vector field is not localized on a brane, which is the well-known
fact \cite{Pomarol, Bajc}. 
For instance, such a non-localization of the vector field occurs
in the Randall-Sundrum model in five dimensions where an additional
localization mechanism, for instance, the Dvali-Shifman mechanism
\cite{Dvali}, must be invoked. 
Nevertheless, and despite the fact that the Dvali-Shifman mechanism
works well for the trapping of vector field, I believe that using
such a particular mechanism, where the peculiar features of the
non-abelian gauge dynamics, those are, the confining property of 
gluodynamics and the dynamical mass gap generation are fully utilized,
as a way to resolve the questions associated with the localization
of bulk fields is not so attractive since the trapping should be 
achieved in terms of the universal interaction, that is, gravity.   
Indeed, in six dimensions, the unique nonsingular solution with
a warp factor to Einstein's equations is known to be $A = c_1 |r|, 
B = c_2 |r|$ with positive constants $c_1$ and $c_2$ \cite{Gregory, Oda3}, 
and then 
the localization condition takes the form $(D_1 - 4)c_1 + D_2 c_2 > 0$
with $D_2$ being fixed to be 1. This time, when $D_1 = 4$, it turns out 
that the vector field is surely localized on a 3-brane by only the 
gravitational interaction without introducing an additional mechanism
\cite{Oda3}. 
This observation appears to encourage us to study further six dimensional
theories in future.

Let us consider spin 3/2 gravitino field \cite{Oda5}. 
The trapping of the gravitino
might be automatic when the graviton is trapped and the theory is
supersymmetrized  (since the gravitino is anyway a superpartner of the
graviton \footnote{This observation is implicitly used in the construction of
a supersymmetric Randall-Sundrum model \cite{Pomarol2}.}), 
but at the present time of writing this article we 
do not have such a supersymmetric theory except in the five dimensions,
so it is valuable to pursue the issue along the same line of arguments as
above. The action for spin 3/2 bulk gravitino is given by the 
Rarita-Schwinger action
\begin{eqnarray}
S_{3/2} = \int d^Dx \sqrt{-g} \bar{\Psi}_M i \Gamma^{[M} \Gamma^N
\Gamma^{R]} (D_N + \delta_N^r \Gamma^r m \varepsilon(r)) \Psi_R, 
\label{10}
\end{eqnarray}
where $D_M \Psi_N = \partial_M \Psi_N - \Gamma^R_{MN} \Psi_R + 
\frac{1}{4} \omega_M^{AB} \gamma_{AB} \Psi_N$ and the square bracket
denotes the anti-symmetrization with weight 1. With the ansatz
$\Psi_r = \Psi_m = 0$, the equations of motion are reduced to
the form 
\begin{eqnarray}
g^{\mu\nu} \left[\Gamma^r (\partial_r - \frac{D_1 - 2}{4} A' 
- \frac{D_2}{4} B') + \Gamma^n (\partial_n + \tilde{\omega}_n)
+ m \varepsilon(r)\right] \Psi_\nu = 0, 
\label{11}
\end{eqnarray}
where we have used equations $\gamma^\mu \Psi_\mu = \hat{D}^\mu \Psi_\mu
= \gamma^{[\mu} \gamma^\nu \gamma^{\rho]} \hat{D}_\nu \Psi_\rho = 0$.
If we look for a solution with the form $\Psi_\mu(x^M) = \psi_\mu (x^\mu)
u(r) \chi(y^m)$, we can get the massless zero-mode solution 
$u(r) = u_0 e^{\frac{D_1-2}{4}A + \frac{D_2}{4}B - m \varepsilon(r) r}$
where we have used $\Gamma^n(\partial_n + \tilde{\omega}_n) \chi =0$
and the chirality condition $\Gamma^r \psi_\mu = \psi_\mu$.
Substituting this solution into the action, we arrive at the following
expression
\begin{eqnarray}
S_{3/2} = u_0^2 \int d^{D_2}y \sqrt{\tilde{g}} |\chi(y)|^2
\int_{-\infty}^\infty dr e^{\frac{1}{2}A - 2 m \varepsilon(r) r}
\int d^{D_1}x \sqrt{-\hat{g}} \bar{\psi}_\mu i \gamma^{[\mu} 
\gamma^\nu \gamma^{\rho]} \hat{D}_\nu \psi_\rho. 
\label{12}
\end{eqnarray}
Again the condition for the localization of the gravitino on a
brane requires the integral over $r$ to take a finite value.
At this stage, it is of interest to notice that the condition has
the same form as in spin 1/2 spinor field discussed before, so just
when spin 1/2 fermion is localized, spin 3/2 gravitino is also 
localized on a brane.

Finally, let us deal with spin 2 graviton. It is well known that
even if free field actions for any higher spin do indeed exist,
we cannot construct interacting actions for more than spin 2 at
least within the framework of a local field theory. Thus, it
is now sufficient to consider the remaining spin 2 case for the
study of localization of the whole spin fields. We shall consider
the metric fluctuations: 
\begin{eqnarray}
ds^2 = e^{-A(r)}(\eta_{\mu\nu} + h_{\mu\nu}) dx^\mu dx^\nu + dr^2
+ e^{-B(r)}\tilde{g}_{mn}(y^l) dy^m dy^n, 
\label{13}
\end{eqnarray}
with $\eta_{\mu\nu}$ being the flat Minkowski metric.
Then the equations of motion for the fluctuations $h_{\mu\nu}$
are found to be
\begin{eqnarray}
\frac{1}{\sqrt{-g}} \partial_M (\sqrt{-g} g^{MN} \partial_N h_{\mu\nu})
= 0.  
\label{14}
\end{eqnarray}
It turns out that these equations are equivalent to the equation of
motion of a scalar field if we replace $h_{\mu\nu}$ with $\phi$.
This fact is reasonable since it is known that transverse traceless 
graviton modes in general obey the equation of a massless scalar in 
a curved background. 
Accordingly, we expect that the condition for localization of spin
2 graviton field might be equivalent to that of spin 0 scalar field.
This is in fact the case as shown in what follows. It is easy to 
show that the equations of motion have the massless zero-mode solution
$h_{\mu\nu}(x^M) = \varphi_{\mu\nu}(x^\mu) u_0 \chi(y^m)$ where 
$\eta^{\mu\nu} \partial_\mu \partial_\nu \varphi_{\lambda\rho} = 0$. 
Then the substitution of this solution into the Einstein-Hilbert action
yields
\begin{eqnarray}
S_2 \sim u_0^2 \int d^{D_2}y \sqrt{\tilde{g}} \chi^2(y)
\int_{-\infty}^\infty dr e^{\frac{2-D_1}{2}A - \frac{D_2}{2}B}
\int d^{D_1}x \partial^\rho \varphi^{\mu\nu} \partial_\rho 
\varphi_{\mu\nu} + \cdots. 
\label{15}
\end{eqnarray}
Thus the localization condition, which is connected with the second
integral over $r$, is the same between the graviton and the scalar.
Therefore, whenever the scalar field is localized, the graviton is so.

In conclusion, in this review article we have presented a complete
analysis of localization of all bulk fields on a brane from the
viewpoint of a local field theory. Our presentation is quite general
in the sense that we have taken account of a rather general metric
ansatz, so it would be useful to apply the analysis at hand to a more
concrete problem. As an example, we have presented the results of 
localization for a warp geometry. Since the warp geometry is now under 
active investigations in a brane world, we would summarize the results
for the warp geometry obtained in this article. We have shown that 
spin 0 and 2 fields are naturally localized on a brane. Spin 1 vector
field is localized in six dimensions, while it is not localized in the
other dimensions so we need to invoke an additional mechanism such as
the Dvali-Shifman mechanism. Spin 1/2 and 3/2 fermionic fields are
localized on a brane if they have a bulk mass term with a 'kink'
profile. It is worthwhile to stress that the warp geometry in six
dimensions is of the most interest in that all the bulk fields are 
trapped by only universal gravitational interaction. A direction for future
study would be to construct a more realistic, phenomenological model based 
on our results.

\begin{flushleft}
{\bf Acknowledgement}
\end{flushleft}
We wish to thank T. Gherghetta and A. Pomarol for useful comments.


\end{document}